\documentclass[12pt]{article}

\usepackage{amsthm}
\usepackage{amsmath}
\usepackage{amssymb}
\usepackage{enumerate}
\usepackage[dvipsnames]{xcolor}

\usepackage[letterpaper]{geometry}
\usepackage[bookmarks=true,colorlinks=true,linkcolor=black,citecolor=black,menucolor=black,urlcolor=black]{hyperref}
\usepackage{authblk}

\usepackage{graphicx}

\usepackage{longtable}	

\definecolor{darkgray}{rgb}{0.4,0.4,0.4} 

\newif\ifnotes
\makeatletter
\newcommand{\note}[1]{\@bsphack\ifnotes{
        \vspace{2mm} \\
        \textcolor{teal}{ \textbf{Note.} {#1} }
        \vspace{2mm} \\
    } \fi\@esphack}
\makeatother

\numberwithin{equation}{section}

\newcommand{\N}{\mathbb{N}}

\newcommand{\E}{\mathbb{E}}

\def\1{{\rm l}\hskip -0.21truecm 1}

\title{{\huge\textbf{Time-dependent non-homogeneous stochastic epidemic model of SIR type}}}

\author[a]{Mireia Besalú}
\author[b]{Giulia Binotto}
\affil[a]{\small{Departament de Genètica, Microbiologia i Estadística, Universitat de Barcelona}}
\affil[b]{\small{Department de Matemàtiques, Universitat Autònoma de Barcelona}}

\date{}

\begin{document}

\maketitle

\begin{abstract}
	To better describe the spread of a disease, we extend a discrete time stochastic SIR-type epidemic model of Tuckwell and Williams. We assume the dependence on time of the number of daily encounters and include a parameter to represent a possible quarantine of the infectious individuals.
	We provide an analytic description of this Markovian model and investigate its dynamics. Both a diffusion approximation and the basic reproduction number are derived. Through several simulations, we show how the evolution of a disease is affected by the distribution of the number of daily encounters and its dependence on time.
\end{abstract}

\section{Introduction}

The interest of mathematical models to describe the spread of a disease has a long history. It dates back to eighteenth-century study of Bernoulli on smallpox inoculation \cite{Bernoulli1760}.
Both deterministic and stochastic models have been considered and many factors have been taken into account: infectious agents, mode of transmission, latent periods, temporary or partial immunity, quarantine periods, etc. (see, for example, \cite{Bailey1975, AndersonEA1991, DiekmannEA2013}).
The main advantage of the deterministic approach lies in its more manageable (even if not necessarily simple) analysis. However, the most natural way to describe the spread of a disease is stochastic. This is due to several facts.
On one hand, some phenomena are genuinely stochastic or present random features.
On the other hand, the infective agent is introduced into the population only through a few hosts. Deterministic models cannot capture this component, given that they apply only when a small fraction (not a small number) of the large population is infected. To solve this contrast, demographic stochasticity needs to be taken into account (see \cite{DiekmannEA2013} for an exhaustive explanation).

An important role in mathematical epidemiology is played by the class of models that divide the population into different categories and study the changes of state between them. One of the most classical are the SIR models, where the letters stand for Susceptible, Infectious and Removed, respectively.
The first SIR stochastic model was proposed by McKendrick \cite{McKendrick1926} as a continuous-time version of the deterministic model of Kermack and McKendrick \cite{KermackEA1927}, although bigger attention was given to the chain-binomial model of Reed and Frost (later published by Abbey \cite{Abbey1952}).
From that moment on, many other texts and models have been developed.

Stochastic models can be divided into three major categories: discrete time models, continuous time Markov chain models and diffusion models.
The literature that deals with these three categories is extensive.
Just to name a few, Tuckwell and Williams \cite{TuckwellEA2007} consider a simple discrete time Markovian model in which the total population is constant and individuals meet a random number of other individuals at each time step.
Ferrante et al. \cite{FerranteEA2016} generalize this model adding two new classes. Their results are more suitable for diseases with an initial latency period and the presence of asymptomatic individuals.
On the other hand, Gray et al. \cite{GrayEA2011} directly propose a system of equations to model the dynamics of the population.
Discrete time models are mathematically less complex than those employed in the other two categories. However, they grant a simplified vision of the evolution of a disease which highlights its main characteristics. They also provide a good guide to construct more complex models.
Besides, generally they do not present specific constrains or assumptions that could be in contrast with empirical evidence, as do, for example,  continuous time Markov chain models.

In the present paper, we consider the SIR-type model proposed by Tuckwell and Williams \cite{TuckwellEA2007}. We expand their results by weakening the homogeneity in time, that is, assuming the dependence on time in some parameters, and by including factors that have not been taken into account.
We deal with a discrete time, discrete state space stochastic model built on a generation basis. Time unit is of one-day length. The population is assumed to be closed, homogeneous and homogeneously mixing, and it is divided in the three classes defined above.
We assume that every day any susceptible meets a certain number of different individuals. If one of them is infective, the disease is transmitted with a given probability. While in \cite{TuckwellEA2007} the number of daily encounters is time-homogeneous, in our case it can change at any time step. Moreover, on occasion they fix this quantity for simplicity, while we maintain it random all over the paper. Any individual remains infectious for a fixed number of consecutive epochs. After this period, he or she recovers and becomes immune.
We also add a parameter to the model to consider the case of a possible quarantine for infectious individuals.

We describe this type of model analytically using the methodologies of \cite{TuckwellEA2007} and \cite{FerranteEA2016}.
We derive an explicit structure for the underlying discrete time Markov chain and deduce the probability that a susceptible will become infected at the following epoch. This is the content of Section \ref{section.description}.
Section \ref{section.calculus} is dedicated to describe the dynamics of the model. Two possible situations are taken into account: when the duration of the disease is constant and when infectious individuals remain infectious throughout their life. For both cases, we describe the distribution of new infecteds and derive a diffusion approximation and the basic reproduction number $R_0$. This is the expected number of secondary cases produced by an infectious individual during its infectious period in a virgin population. This value is used to measure the potential for transmission of a disease.
Section \ref{section.simulations} contains some simulations of the SIR-type model we have described. After fixing some parameters, we compare the evolution of the disease for different distributions of the number of daily contacts. We see how the different behaviors and the dependence on time of this parameter affect the course of the disease.

Throughout the paper, we use the term \textit{infective} to refer to an individual who has contracted the disease and is infectious (he or she belongs to class I), while we connote with \textit{infected} an individual who is either infectious or removed, i.e., no longer susceptible (he or she belongs to class I or R).

\section{Description of the model} \label{section.description}

This section is dedicated to describe the basic properties of the SIR model we consider. It is an extension of the one proposed by Tuckwell and Williams \cite{TuckwellEA2007}. The main difference regards the number of daily contacts. While Tuckwell and Williams assume that this quantity is time-homogeneous, in our case it can change at any time step. Moreover, they assume it is a random variable, but they fix it to be a constant in many occasions. On the contrary, we maintain the random character of this parameter throughout all the paper.
We also add a parameter to the model to consider the case that the infective individuals adopte some kind of quarantine. This can be a perfect quarantine (there are no contacts) to no quarantine at all (the infective individuals have the same contacts as a susceptible one).
\bigskip
We assume:
\begin{enumerate}
	\item Total population size: It is fixed at $n$.
	\item Time: Time is discrete. On epidemic the natural unit for the duration of an epoch is one day, although in some applications the time step is bigger (see, for example, \cite{TsutsuiEA2003}).
	\item Definition of a sick individual: Given any individual $i$, with $i=1,\dots,n$, we define a stochastic process $Y^i=\{Y^i(t), t=0,1,2,\dots\}$ such that
		\[ Y^i(t) = \left\{\begin{array}{ll}
			1 & \text{if the individual $i$ is infective (and infectious) at time $t$} \\
			0 & \text{otherwise.}
		\end{array}\right. \]
	Then the total number of infective and hence infectious individuals at time $t$ is
		\[ Y(t) = \sum_{i=1}^n Y^i(t). \]
	\item Daily encounters: Over $(t,t+1]$ each individual $i$ meets a random number $N^i(t)$ of other individuals. For all $i$ and $t$, the variables $N^i(t)$ are mutually independent and independent of the state of the population.
	Furthermore, for all $i$ and $t$, $N^i(t)$ can take only a finite number of values $n_k$, each with probability $p_k(t)$. The values are fixed, that is, they depend neither on individuals nor on time, while its probability varies with respect to time, but not to individuals. This is, for $k=1,\dots,m$,
		\[ P(N^i(t)=n_k) = p_k(t) \qquad\forall i,t \]
	with $\sum_{k=1}^m p_k(t)=1$ for all $t$. Observe that the set of all possible values of $N^i(t)$, that we define $M=\{n_1,\dots,n_m\}$, is a ordered subset of $\N_{\geq0}$.
	As we consider models where, for a fixed $t$, the distribution of the number of daily encounters is the same for all individuals, throughout the paper we will use the simplified notation $N(t)$ instead of $N^i(t)$, when appropriate.
	\item Duration of the disease: Any individual remains infectious for $r$ consecutive epochs, where $r$ is a positive integer. After this period, the individual recovers and becomes immune. The case without recovery, that is when $r=\infty$, is also considered.
	\item Contagion probability: If an individual who has never been diseased up to and including time $t$ encounters an individual in $(t,t+1]$ who is diseased at time $t$, then independently of the results of other encounters, this encounter results in transmission of the disease with probability $p$. Thus, such individual is infected at epoch $t+1$.
	\item Encounter probability: The population is homogeneously distributed, this means that, given $Y(t)=y$, the probability that a randomly chosen individual is infectious at time $t$ is given by $y/n$. We will multiply this probability by a constant $\lambda\in[0,1]$ to characterise a possible quarantine of the infectious individuals. If $\lambda=0$ the infectious individuals do a rigorous quarantine and for values of $\lambda$ close to $1$ they will do a mild or non-existent quarantine. 
\end{enumerate}

As mentioned before, the variables $N^i(t)$ are independent for all $i$ and $t$. Assuming that, for a fixed $t$, they are also identically distributed, the model can be seen as an $(r+1)$-dimensional Markov chain.
Observe that, even if $N^i(t)$ are not identically distributed with respect to time, this model is still a Markov chain, but it is not homogeneous. This means that the Markov property is retained but the transition probabilities may depend on time.

Indeed, let
\begin{itemize}
	\item $Y_\ell(t)$ be the number of individuals who are infective at time $t$ and have been infective for exactly $\ell$ time units, with $\ell=0,1,\dots,r-1$;
	\item $X(t)$ be the number of susceptible individuals at time $t$;
	\item $Z(t)$ be the number of individuals who were previously infective and are recovered at time $t$.
\end{itemize}

We assume that all of the individuals who are infective at $t=0$ have just become infected so that $Y(0)=Y_0(0)$ and $Y_\ell(0)=0$ for $\ell=1,\dots,r-1$. Also, there are no recovered individuals at $t=0$, so that $Z(0)=0$, and the population is made up only with susceptible and just infected individuals, so $Y_0(0)+X(0)=n$.

\bigskip
Regardless of the initial conditions, the process
	\[ M(t) = \left( X(t),Y_0(t),Y_1(t),\dots,Y_{r-1}(t) \right) \qquad t=0,1,2,\dots \]
is a Markov chain with state space
	\[ S(n,r) = \left\{ (x,y_0,\dots,y_{r-1})\in\N^{r+1} \,\text{ s.t }\, x+\sum_{\ell=0}^{r-1}y_\ell\leq n  \right\}. \]
\note{\hypertarget{note_MCNotHomog}{$M(t)$ is non-homogeneous (see the transition probabilities). Moreover, if $N^i(t)$ depends on $i$, then $M(t)$ is not a Markov chain.}}
	
The cardinality of $S(n,r)$ is $\binom{n+r+1}{n}$.
\note{The problem is equivalent to having $n$ balls and $r+1$ separators. In total, we have $n+r+1$ objects and we want to place the $r+1$ separators, i.e., we want to choose the positions where to put the $r+1$ separators between the $n+r+1$ possible ones. There are $\binom{n+r+1}{r+1}=\binom{n+r+1}{n}$ way to do this.}
Observe that the values of $Z(t)$ are determined if all the components of $M(t)$ are known.

The total number of infectives at time $t$ is given by
	\[ Y(t) = \sum_{\ell=0}^{r-1} Y_\ell(t), \]
so the set $\left( X(t),Y(t),Z(t) \right)$ gives the traditional SIR description.

In addition to the process $Y^i=\{Y^i(t),t=0,1,2,\dots\}$, we can define in a similar manner the process $X^i=\{X^i(t),t=0,1,2,\dots\}$, for $i=1,\dots,n$, which indicates whether individual $i$ is susceptible or not, and the variable $Z^i(t)=1-X^i(t)-Y^i(t)$ which indicates if the individual $i$ has recovered from the disease  and is no longer infectious.
For all $t$, we get
	\[ X(t)=\sum_{i=1}^n X^i(t), \quad Y(t)=\sum_{i=1}^n Y^i(t) \quad\text{and}\quad Z(t)=\sum_{i=1}^n Z^i(t). \]

\bigskip
Furthermore, even if the random variables $N^i(t)$ are not identically distributed with respect to time, for $i=1,\dots,n$, we can consider the processes $Y^i_0,Y^i_1,\dots,Y^i_{r-1}$ where
	\[ Y^i_\ell(t) = \left\{\begin{array}{ll}
		1 & \text{if the individual $i$ at time $t$ is infective for $\ell$ days} \\
		0 & \text{otherwise.}
	\end{array}\right. \]
Hence,
	\[ Y^i(t) = \sum_{\ell=0}^{r-1} Y^i_\ell(t) \quad\text{and}\quad Y_\ell(t) = \sum_{i=1}^n Y^i_\ell(t). \]	
Then, we can consider another Markov chain
	\[ \widetilde{M}(t) = \left( X^i(t),Y^i_0(t),Y^i_1(t),\dots,Y^i_{r-1}(t), \,i=1,\dots,n \right) \qquad t=0,1,2,\dots \]
with state space
\begin{eqnarray*}
	\widetilde{S}(n,r) = \bigg\{ (x^1,y^1_0,\dots,y^1_{r-1},\dots,x^n,y^n_0,\dots,y^n_{r-1})\in\{0,1\}^{n(r+1)} \,\text{ s.t } \\
	\alpha^i=x^i+\sum_{\ell=0}^{r-1}y^i_\ell\leq 1 \text{ for } i=1,\dots,n \,\text{ and } \sum_{i=1}^n \alpha^i\leq n \bigg\}.
\end{eqnarray*}
\note{$\widetilde{M}(t)$ is not homogeneous (see previous \hyperlink{note_MCNotHomog}{note}). \textcolor{Green}{Incorporar aquesta Note a l'article?}}
The cardinality of $\widetilde{S}(n,r)$ is equal to $(r+2)^n$.
Even if the cardinality is bigger than the one of $S(n,r)$, this Markovian model is more simple and more suitable for simulation purposes.

\subsection{Transition probabilities}

For a fixed individual $i$, consider the process
	\[ \widetilde{M}^i(t) = \left( X^i(t),Y^i_0(t),Y^i_1(t),\dots,Y^i_{r-1}(t) \right) \qquad t=0,1,2,\dots \]
If one of the variables $Y^i_0(t),Y^i_1(t),\dots,Y^i_{r-1}(t)$ is equal to 1, then the process at times bigger than $t$ is certain since the transition in this case is sure. The only interesting case is when the individual is susceptible at time $t$, that is, $X^i(t)=1$.

Let calculate the probability that an individual $i$ susceptible at $t$ becomes infected for the first time at $t+1$. This probability depends on the total number $Y(t)=y$ of diseased individuals together with the probability $p$ of transmission per contact and the parameter $\lambda\in[0,1]$ characterising a possible quarantine. It is calculated taking into account all possible values of $N^i(t)$ and its probabilities.
Recall that the variables $N^i(t)$ are identically distributed for all individuals, but not with respect to time, so we will use the notion $N(t)$ instead.
Assuming $n$ is much greater than $N(t)$, so that the binomial approximation may be used, the probability of meeting exactly $j$ infectives if $N(t)=n_k$ individuals are met is
\begin{equation}\label{eq_ProbMeet}
	p^i_j(y,n_k;n) \approx {n_k\choose j} \left( \frac{\lambda y}{n-1} \right)^j \left( 1-\frac{\lambda y}{n-1} \right)^{n_k-j}
\end{equation}
when $y<n-1$, while $p^i_{n_k}(y,n_k;n)=1$ and $p^i_j(y,n_k;n)=0$ for $j=0,\dots,n_k-1$ when $y=n-1$.
The probability $p_j$ of becoming infected if $j$ infectives are met is
\begin{equation}
	p_j = 1-(1-p)^j.
\end{equation}
Then, using \eqref{eq_ProbMeet} as an equality, the probability that an individual $i$ susceptible at $t$ becomes infected for the first time at $t+1$ is
\begin{eqnarray}\label{eq_p(t,y)}
	p(t,y) &=& P(Y^i_0(t+1)=1|X^i(t)=1,Y(t)=y) \nonumber \\
	&=& \sum_{k=1}^m P(Y^i_0(t+1)=1|X^i(t)=1,Y(t)=y,N(t)=n_k) \,P(N(t)=n_k) \nonumber \\
	&=& \sum_{k=1}^m \sum_{j=0}^{n_k} p^i_j(y,n_k;n) \,p_j \,p_k(t) \nonumber \\
	&=& \sum_{k=1}^m \sum_{j=0}^{n_k} {n_k\choose j} \left( \frac{\lambda y}{n-1} \right)^j \left( 1-\frac{\lambda y}{n-1} \right)^{n_k-j} \,[1-(1-p)^j] \,p_k(t) \nonumber \\
	&=& \sum_{k=1}^m p_k(t) \sum_{j=0}^{n_k} {n_k\choose j} \left( \frac{\lambda y}{n-1} \right)^j \left( 1-\frac{\lambda y}{n-1} \right)^{n_k-j} \nonumber \\
	&& \hspace{10mm} + \sum_{k=1}^m p_k(t) \sum_{j=0}^{n_k} {n_k\choose j} \left( \frac{(1-p)\lambda y}{n-1} \right)^j \left( 1-\frac{\lambda y}{n-1} \right)^{n_k-j} \nonumber \\
	&=& 1 - \sum_{k=1}^m \left( 1-\frac{p\lambda y}{n-1} \right)^{n_k} \,p_k(t).
\end{eqnarray}
when $y<n-1$, while $p(t,n-1) = 1 - \sum_{k=1}^m (1-p)^{n_k} \,p_k(t)$.

Note that, as commonly used, this model contains a simplification regard the meeting between individuals: the meeting relationship is not symmetric because if the group randomly chosen to meet individual $i$ contains individual $j$, the group chosen to meet individual $j$ does not necessarily contain individual $i$.

\section{Calculus on the model} \label{section.calculus}

After presenting our model, we are interested in describe its dynamics. We study the distribution of new infective individuals and the basic reproduction number. Then, we prove that our model can approximate a diffusion process.
We consider two cases. The first is a general one, where any individual remains infectious for $r<\infty$ consecutive epochs. Here $r$ is a fixed positive constant. We refer to this model as the one with recovery. The second is the particular case without recovery, that is when $r=\infty$.

\subsection{The model with recovery ($r<\infty$)}

Recall that the processes $X(t)$, $Y(t)$ and $Z(t)$ represents the number of susceptible, infective and (previously infected and) recovered individuals at time $t$, respectively. As their sum is fix and correspond to the size of the population, to study the number of new infective individuals it is sufficient to know just two of these quantities.
Let define $V(t)$ as the individuals non suspectibles at time $t$, that is $V(t)=Y(t)+Z(t)$. It is useful to observe that the processes $Y(t)$ and $Z(t)$ can be expressed in terms of $V(t)$:
    \[ Y(t)=V(t)-V(t-r) \quad\text{and}\quad Z(t)=V(t-r). \] 
The number of new infectives at time $t+1$ is given by $V(t+1)-V(t)$, so the total number of infectives at time $t+1$ is given by the number of new infectives and the number of infectives at $t$ who have not yet recovered at $t+1$.

Then, following the previous computations we have 
\begin{eqnarray*}
	&& P(V(t+1)=w+y+z| Y(t)=y, Z(t)=z)\\[4mm]
	&& \hspace{3cm} = P(V(t+1)=w+y+z| V(t)-V(t-r)=y, V(t-r)=z)\\[4mm]
	&& \hspace{3cm} = {n-y-z \choose w} p(t,y)^w (1-p(t,y))^{n-y-z-w}.
\end{eqnarray*}
Therefore, the distribution of the increment in the number of infectives follows a binomial law:
	\[ V(t+1)-V(t)|Y(t)=y,Z(t)=z \sim Binom(n-y-z,p(t,y)). \]
When $y<n-1$, its mean and variance are
\begin{equation*}
	\E[V(t+1)-V(t)|Y(t)=y,Z(t)=z] = (n-y-z) \left[ 1 - \sum_{k=1}^m \left( 1-\frac{p\lambda y}{n-1} \right)^{n_k} \,p_k(t) \right]
\end{equation*}
and
\begin{eqnarray*}
	&& \text{Var}[V(t+1)-V(t)|Y(t)=y,Z(t)=z] \\
	&& \hspace{10mm} = (n-y-z) \left[ 1 - \sum_{k=1}^m \left( 1-\frac{p\lambda y}{n-1} \right)^{n_k} \,p_k(t) \right] \left[ \sum_{k=1}^m \left( 1-\frac{p\lambda y}{n-1} \right)^{n_k} \,p_k(t) \right].
\end{eqnarray*}
In the extreme case when $y=n-1$, $z$ can take only two possible values. If $z=1$, it means that there are no more susceptible individuals in the population and the number of new infectives is constantly zero. If $z=0$, all individuals are infective except one. Then, the distribution of the increment in the number of infectives corresponds to the infection of the unique susceptible and follows a Bernoulli law with parameter $p(t,n-1)$. Its means and variance are easily deduced.

\subsubsection{The basic reproduction number $R_0$}

The basic reproduction number $R_0$ is the expected number of secondary cases produced by an infective individual during its period of infectiousness in a population where all individuals are susceptible to infection. $R_0$ excludes new cases produced by the secondary cases.
The basic reproduction number is used to measure the transmission potential of a disease, as an epidemic occurs in a susceptible population only if $R_0>1$.

Our idea consists in estimating the basic reproduction number observing the role of the infective individual and determining how many people he or she infects at any time step. A different approach has been used, for example, in \cite{FerranteEA2016}, where the estimation is based on the probability of a susceptible individual to be infected by the tagged one.
In this section, we first consider the case without quarantine and then deduce the basic reproduction number when a quarantine for infectious individuals has been established.

\bigskip

In accordance with the definition of the basic reproduction number, we assume that at time $t=0$ the number of susceptible individuals is $X(0)=n-1$ and that the number of infected ones is $Y(0)=1$. Clearly, $Z(0)=0$.

Suppose first that there is no quarantine, that is $\lambda=1$.
We denote by $I(t)$ the number of individuals infected by the our tagged individual during only the $t$-th period and by $S(t)$ the number of susceptible individuals met by one person at time $t$. We assume that a susceptible individual, met and infected by another (for example, the tagged one) at time $t$, is considered infective since time $t+1$.
Since our tagged individual remains infectious for $r$ consecutive days (from $t=0$ to $t=r-1$), the basic reproduction number is given by
	\[ R_0 = \sum_{t=1}^r \E[I(t)]. \]
For all $t$, $I(t)$ depends on the number $S(t-1)$ of susceptible individuals met by the tagged one at time $t-1$ and the probability of transmission $p$.

Since at time $t=0$ all individuals except the tagged one are susceptible, $S(0)=N(0)$ and the distribution of $I(1)$ is quite simple to determine:
	\[ I(1)|N(0)=n_k \sim Binom(n_k,p). \]
Consequently,
	\[ \E[I(1)] = \E \left[ \E[I(1)|N(0)] \right] = p \,\E[N(0)]. \]

For all $t\in\{1,\dots,r-1\}$, the variable $S(t)$ may not coincide with $N(t)$ and the distribution of $I(t+1)$ is given by
	\[ I(t+1)|S(t)=l \sim Binom(l,p). \]
Moreover, for all $t\in\{1,\dots,r-1\}$, $S(t)$ depends on the number of individuals met by the tagged one and on the number of infective in the total population. Observe that during this period the number of removed individual is always zero, since our tagged one will be the first removed individual at time $t=r$.
$S(t)$ follows an hypergeometric distribution:
	\[ S(t)|N(t)=n_k,Y(t)=y \sim HGeom(n-1,n-y,n_k). \]
Observe that we have the following parameters: $n-1$ is the size of the population without counting the tagged individual; $n-y$ represents the number of suceptible individuals in the population; and $n_k$ is the number of daily contact of one individual.
As $n$ is supposed to be much greater than $N(t)$, we use the binomial approximation
\begin{equation} \label{distrib.St}
    S(t)|N(t)=n_k,Y(t)=y \approx Binom(n_k,p(y))
\end{equation}
where
	\[ p(y) = 1-\frac{y-1}{n-1} \]
is the proportion of susceptible individuals in the population (without counting the tagged individual).

Then, for all $t\in\{1,\dots,r-1\}$, we have
\begin{eqnarray*}
	E[I(t+1)] &=& \E \left[ \E[I(t+1)|S(t)] \right] = p \,\E[S(t)] \\
	&=& p \,\E \left[ \E[S(t)|N(t),Y(t)] \right] \\
	&=& p \,\E[N(t) \cdot p(Y(t))] \\
	&=& p \,\E[N(t)] \left( 1-\frac{\E[Y(t)]-1}{n-1} \right).
\end{eqnarray*}
In the last step, we use the assumption that the number of daily contacts does not depend on the state of the population, which means that $N(t)$ is independent of $Y(t)$.
Now, using an induction argument, we show that
\begin{equation} \label{eq.mean.pY}
    \E[p(Y(t))] = 1-\frac{\E[Y(t)]-1}{n-1} = 1+O\left(\frac1{n-1}\right).
\end{equation}
\note{Recall that $f(n)=O(g(n))$ if $\exists\, n_0, C>0$ s.t. $|f(n)|\leq C|g(n)|$ for all $n\geq n_0$.}
First, observe that at time $t=1$ the expected number of infectives is given by the tagged individual and the ones he or she has infected:
\begin{equation} \label{eq.mean.Y1}
    \E[Y(1)] = 1+\E[I(1)] = 1+p\,\E[N(0)].
\end{equation}
Thus,
\begin{equation} \label{eq.mean.pY1}
    \E[p(Y(1))] = 1-\frac{p\,\E[N(0)]}{n-1} = 1+O\left(\frac1{n-1}\right).
\end{equation}
Recall that the number of daily contacts is assumed to be much smaller than the size of the population.

To better understand how $\E[p(Y(t))]$ behaves, we show that \eqref{eq.mean.pY} is fulfilled also for $t=2$. The idea is that the expected number of infectives is given by the number of infectives at $t=1$ and by the new infected to which they gave rise, and that the number of individuals infected by one single person depends on the probability of transmission $p$ and the number $S(1)$ of susceptibles met in the previous time step.
This is
    \[ \E[Y(2)] = \E[Y(1)] + \E[Y(1)] \cdot p \,\E[S(1)] = \E[Y(1)] \left( 1+p \,\E[S(1)] \right). \]
From \eqref{distrib.St} and \eqref{eq.mean.pY1}, we have
    \[ \E[S(1)] = \E\big[\E[S(1)|N(1)]\big] = \E[N(1)] \cdot \E[p(Y(1))] = \E[N(1)] \left( 1+O\left(\frac1{n-1}\right) \right). \]
Then, from \eqref{eq.mean.Y1}
\begin{eqnarray*}
    \E[Y(2)] &=& \left( 1+p\,\E[N(0)] \right) \left( 1+p\,\E[N(1)]+O\left(\frac1{n-1}\right) \right) \\
    &=& 1+p\,\E[N(0)]+p\,\E[N(1)]+p^2\,\E[N(0)]\,\E[N(1)]+O\left(\frac1{n-1}\right)
\end{eqnarray*}
and
\begin{eqnarray*}
    \E[p(Y(2))] &=& 1-\frac{1+p\,\E[N(0)]+p\,\E[N(1)]+p^2\,\E[N(0)]\,\E[N(1)]}{n-1}+O\left(\frac1{(n-1)^2}\right) \\
    &=& 1+O\left(\frac1{n-1}\right).
\end{eqnarray*}
Now, suppose that \eqref{eq.mean.pY} holds for $t$ and prove that it is also true for $t+1$. Following the same ideas used for $t=2$, we have
    \[ \E[Y(t+1)] = \E[Y(t)] \left( 1+p\,\E[S(t)] \right) \]
and
    \[ \E[S(t)] = \E[N(t)] \left( 1+O\left(\frac1{n-1}\right) \right). \]
Then,
\begin{eqnarray*}
    \E[p(Y(t+1))] &=& 1-\frac{\E[Y(t)]-1}{n-1} \left( 1+p\,\E[N(t)]+O\left(\frac1{n-1}\right) \right) \\
    &=& 1+O\left(\frac1{n-1}\right) \cdot \left( 1+p\,\E[N(t)]+O\left(\frac1{n-1}\right) \right) \\
    &=& 1+O\left(\frac1{n-1}\right).
\end{eqnarray*}
Then, \eqref{eq.mean.pY} is proved. Observe that, following the same ideas used for $t=2$, it is possible to obtain an explicit expression for $\E[Y(t)]$ as sum of terms of type $p^k\,\E[N(t_1)]\cdots\E[N(t_k)]$ with $k\in\{1,\dots,t\}$ and $t_1,\dots,t_k\in\{0,t-1\}$. Since the notation is complicated and does not contribute significantly to the proof of equation \eqref{eq.mean.pY}, we have decided to omit these calculations.

It follows that, for all $t\in\{1,\dots,r-1\}$,
    \[ E[I(t+1)] = p\,\E[N(t)]+O\left(\frac1{n-1}\right). \]


Finally, we get
    \[ R_0 = \sum_{t=1}^r \E[I(t)] = p \sum_{t=0}^{r-1} \E[N(t)] + O\left(\frac1{n-1}\right). \]

\bigskip
If we assume that there is a quarantine, recall that the probability that a randomly chosen individual is infectious is multiplied by a parameter $\lambda\in[0,1]$. Thinking about the role of this parameter, we see that it also affects the contagiousness of an infective individual, that is also multiplied by the same factor. This means that the basic reproduction number is given by
	\[ R_0 = \lambda\sum_{t=1}^r \E[I(t)]. \]
The same calculations used for the case without quarantine lead to the following expression for $R_0$:
\begin{equation} \label{eq.R0}
    R_0 = \lambda p \sum_{t=0}^{r-1} \E[N(t)] + O\left(\frac1{n-1}\right).
\end{equation}

As the number of susceptibles met by the tagged one is always less or equal to the number of his/her encounters, we can deduce an upper bound for $R_0$:
	\[ R_0 \leq \lambda p \sum_{t=0}^{r-1} \E[N(t)]. \]

From \eqref{eq.R0} it follows that, since $R_0\approx 1$ for $\lambda p \sum_{t=0}^{r-1} \E[N(t)]=1$, this is a threshold for the epidemic. This means that it grows for larger values and dies soon for smaller ones.

\subsubsection{A diffusion approximation}

Following the ideas of  Tuckwell and Williams \cite{TuckwellEA2007}, the study of the mean and variance of the one-step increments of $V$ indicates that for a large population size $n$ and a small probability transition $p$ such that $np\E[N([nt])]$ is of moderate size for all $t$, we can approximate a rescaled version of $V$ by a diffusion process.

More accurately, if we speed up time and rescale the state we can define a process
    \[ \widehat{V}^n(t)=\frac{V([nt])}{n} \qquad \textrm{for all}\;\; t\geq 0 \]
where $[\cdot]$ denotes the greatest integer part. $\widehat{V}^n(t)$ can be interpreted as the fraction of the population that has been infected by the time $[nt]$ in the original time scale of $V$. Then, for $n$ large and $p$ small such that $\theta(t)=np\E[N([nt])]$ is of moderate size for all $t$, we see that with $\Delta t =\frac1n$ and $t=0,\,\frac1n,\,\frac2n,\ldots$
\begin{eqnarray*}
    && \E \left[ \widehat{V}^n(t+\Delta t)-\widehat{V}^n(t) \Big| \widehat{V}^n(t)-\widehat{V}^n\left(t-\frac rn\right)=\hat{y}, \widehat{V}^n\left(t-\frac rn\right)=\hat{z} \right] \\
    && \hspace{5mm} = \frac1n \E \left[ V([nt]+1)-V([nt]) \Big| V([nt])-V([nt]-r)=n\hat{y}, V([nt]-r)=n\hat{z} \right] \\
    && \hspace{5mm} = \frac1n (n-n\hat{y}-n\hat{z}) \left[ 1-\sum_{k=1}^m \left(1-\frac{\lambda np\hat{y}}{n-1}\right)^{n_k} p_k([nt]) \right] \\
    && \hspace{5mm} \approx (1-\hat{y}-\hat{z}) \frac{\lambda np\hat{y}}{n-1} \E[N([nt])] \\
    && \hspace{5mm} \approx \lambda \theta(t) \hat{y}(1-\hat{y}-\hat{z}) \Delta t
\end{eqnarray*}
and
\begin{eqnarray*}
    && \text{Var} \left[ \widehat{V}^n(t+\Delta t)-\widehat{V}^n(t) \Big| \widehat{V}^n(t)-\widehat{V}^n\left(t-\frac rn\right)=\hat{y}, \widehat{V}^n\left(t-\frac rn\right)=\hat{z} \right] \\
    && \hspace{5mm} = \frac1{n^2} \text{Var} \left[ V([nt]+1)-V([nt]) \Big| V([nt])-V([nt]-r)=n\hat{y}, V([nt]-r)=n\hat{z} \right] \\
    && \hspace{5mm} = \frac1{n^2} (n-n\hat{y}-n\hat{z}) \left[ 1-\sum_{k=1}^m \left(1-\frac{\lambda np\hat{y}}{n-1}\right)^{n_k} p_k([nt]) \right] \left[ \sum_{k=1}^m \left(1-\frac{\lambda np\hat{y}}{n-1}\right)^{n_k} p_k([nt]) \right] \\
    && \hspace{5mm} \approx \frac1n (1-\hat{y}-\hat{z}) \left[ \frac{\lambda np\hat{y}}{n-1} \E[N([nt])] \right] \left[ 1-\frac{\lambda np\hat{y}}{n-1} \E[N([nt])] \right] \\
    && \hspace{5mm} \approx \frac1n (1-\hat{y}-\hat{z}) \frac{\lambda np\hat{y}}{n-1} \E[N([nt])] \\
    && \hspace{5mm} \approx \frac{\theta(t)\lambda }{n} \hat{y}(1-\hat{y}-\hat{z}) \Delta t
\end{eqnarray*}
In both calculation we use the approximation $1-(1-x)^a\approx ax$ for small $x$. In the calculation of the variance we also use the approximation $x(1-x)\approx x$ for small $x$.

Moreover, we recall that $\E[N([nt])]$ depends on the values $n_1,n_2,\dots,n_m$ and its probabilities $p_k(t),\, k=1,\ldots,m$. In fact, the time dependence on $\E[N([nt])]$ rely on $p_k(t)\in[0,1],\, k=1,\ldots,m$. So, we can assume for all $n$ and $t$ that $\E[N([nt])]$ is of order of a constant $N$, so
\[\theta(t)=np\E[N([nt])]\approx npN:=\theta.\]
With the previous results and approximation methods for continuous time Markov chains using diffusion processes (similar results to \cite{FerranteEA2016} and \cite{TuckwellEA2007}), we can approximate $\widehat{V}^n$ by a diffusion process $\widehat{V}$ in $[0,1]$ that satisfies the stochastic delay differential equation (SDDE)
\begin{equation} \label{eq:SDDE}
	d\widehat{V}(t)=\lambda\theta\left(\widehat{V}(t)-\widehat{V}(t-\tau)\right)(1-\widehat{V}(t))dt+\left[\frac{\lambda\theta}{n}\left(\widehat{V}(t)-\widehat{V}(t-\tau)\right)(1-\widehat{V}(t))\right]^\frac12 dW(t), 
\end{equation}

where $\tau=\frac{r}{n}$ and $W=\{W(t),t\geq 0\}$ is a standard Brownian motion. This equation has a unique solution for a given initial condition $\widehat{V}_0$, which is a non-decreasing continuous function $\widehat{V}_0:[-\tau,0]\rightarrow[0,1]$ (see \cite{Mohammed1998} for more references in SDDE).
\vskip 5pt
Moreover, if we can assume that the variability is small, then the noise term in \eqref{eq:SDDE} has little effect.  So it seems natural to conjecture that the mean of $\widehat{V}(t)$ can be approximated by $\widehat{m}(t)$, where $\widehat{m}$ satisfies the deterministic equation
	\[ d\widehat{m}(t)=\lambda\theta\left(\widehat{m}(t)-\widehat{m}(t-\tau)\right)\left(1-\widehat{m}(t)\right)dt \]
\note{In the first approximation, $x=\frac{np\hat{y}}{n-1}$ and $a=n_k$. In the second approximation, $x=\frac{np\hat{y}}{n-1} \E[N([nt])]$. As we assume $np\E[N([nt])]$ of moderate size, we can apply both.}
\note{Intuitively (see \cite{RogersEA2000}, Chapter 5.1), I think that it is sufficient suppose $np\E[N([nt])]\approx\sqrt{n}$, with $np\approx\sqrt{n}$ and $\E[N([nt])]\approx C$ constant.}

\subsection{The model without recovery ($r=\infty$)}

Consider now the case in which the infectious individuals remain infectious throughout the course of the epidemic. Such a situation can arise when a disease causing agent has a long life, as with tuberculosis in deer \cite{WahlstromEA1998}, or when life-prolonging drug therapies have been found, as with HIV in humans \cite{NgEA1990}. As recovered individuals are not presented in this case, the model reduces to one of SI type rather than SIR.

Here, only two classes of individuals are present: the susceptibles and the infectives. Their amount at time $t$ is given by the variables $X(t)$ and $Y(t)$, respectively.

The distribution of new infective individuals only depends on the number of infectives of the previous generation. Assuming that at time $t$ there are $Y(t)=y$ infectives, implies that there are $n-y$ susceptibles, since the population size is $n$ and there are no recovered individuals.
Recall that the distribution of daily encounters at time $t$, given by the variables $N(t)$, is the same for all individuals. Moreover, $N(t)$ are independent for all $i$ and $t$. In this case, $\lambda$ parameter has no sense and will be foxed $\lambda=1$ (no quarantine).

To study the distribution of new infectives, we follow the same ideas of the case with recovery, with the difference that here it is not necessary to define a new variable that represents the number of non-susceptible individuals, as this is given by $Y(t)$.
Given $Y(t)=y$, the number of new infectives at $t+1$ follows a binomial distribution whose parameter are the number of susceptibles at $t$ and the probability that an individual susceptible at $t$ becomes infected for the first time at $t+1$. This is
	\[ Y(t+1)-Y(t)|Y(t)=y \sim Binom(n-y,p(t,y)) \]
where
	\[ p(t,y) = 1 - \sum_{k=1}^m \left( 1-\frac{py}{n-1} \right)^{n_k} \,p_k(t) \]
is given by \eqref{eq_p(t,y)}.
This implies that the distribution of the increment in the number of infectives is given by
\begin{eqnarray}
	&& P(Y(t+1)=y+w|Y(t)=y) \nonumber \\
	&& \hspace{10mm} = {n-y\choose w} \left[ 1 - \sum_{k=1}^m \left( 1-\frac{py}{n-1} \right)^{n_k} \,p_k(t) \right]^w \left[ \sum_{k=1}^m \left( 1-\frac{py}{n-1} \right)^{n_k} \,p_k(t) \right]^{n-y-w} \nonumber \\
\end{eqnarray}
for $w=0,\dots,n-y$. Its mean is
\begin{equation*}
	\E[Y(t+1)-Y(t)|Y(t)=y] = (n-y) \left[ 1 - \sum_{k=1}^m \left( 1-\frac{py}{n-1} \right)^{n_k} \,p_k(t) \right]
\end{equation*}
and its variance is
\begin{eqnarray*}
	&& \text{Var}[Y(t+1)-Y(t)|Y(t)=y] \\
	&& \hspace{10mm} = (n-y) \left[ 1 - \sum_{k=1}^m \left( 1-\frac{py}{n-1} \right)^{n_k} \,p_k(t) \right] \left[ \sum_{k=1}^m \left( 1-\frac{py}{n-1} \right)^{n_k} \,p_k(t) \right].
\end{eqnarray*}

\subsubsection{The basic reproduction number $R_0$}

In this section we calculate the value of the basic reproduction number in the case without recovery, that we denote by $R_0^{(\infty)}$.
The computation for the case with recovery can be adapted to this model.

Here, our tagged individual is infectious all his/her life, but we can suppose that he or she lives only for a amount $T$ of days after being infected. Thus, the calculation of the basic reproduction number is the same than the one of the case with recovery, changing the number of terms from $r$ to $T$:
	\[ R_0^{(\infty)} = \sum_{t=1}^T \E[I(t)] = p \sum_{t=0}^{T-1} \E[N(t)] + O\left(\frac1{n-1}\right). \]
Recall that $I(t)$ denotes the number of individuals infected by the our tagged individual during only the $t$-th period and observe that in this case no quarantine is established for infectives individuals.

The considerations about an upper bound for $R_0^{(\infty)}$ and a threshold for the epidemic can be done also in the case without recovery.
Indeed, on one hand, as the number of susceptibles met by the tagged one is always less or equal to the number of his/her encounters, an upper bound for $R_0^{(\infty)}$ is:
		\[ R_0^{(\infty)} \leq p \sum_{t=1}^{T-1} \E[N(t)]. \]
On the other hand, since $R_0^{(\infty)}\approx 1$ for $p \sum_{t=1}^{T-1} \E[N(t)]=1$, this is a threshold for the epidemic.

\subsubsection{A diffusion approximation}

Following the ideas of  Tuckwell and Williams \cite{TuckwellEA2007}, the study of the mean and variance of the one-step increments of $Y$ indicates that for a large population size $n$ and a small probability transition $p$ such that $np\E[N([nt])]$ is of moderate size for all $t$ fixed, we can approximate a rescaled version of $Y$ by a diffusion process.

More accurately, if we speed up time and rescale the state we can define a process
\[\widehat{Y}^n(t)=\frac{Y([nt])}{n}\qquad \textrm{for all}\;\; t\geq 0\]
where $[\cdot]$ denotes the greatest integer part. Then, we can interpret $\widehat{Y}^n(t)$ as the fraction of the population that has been infected by the time $[nt]$ in the original time scale of $Y$. Then, for $n$ large and $p$ small such that $\theta(t)=np\E[N([nt])]$ is of moderate size for all $t$ we see that with $\Delta t =\frac1n$ and $t=0,\,\frac1n,\,\frac2n,\ldots$
\begin{eqnarray*}
&&\hspace{-10pt}\E[\widehat{Y}^n(t+\Delta t)-\widehat{Y}^n(t)| \widehat{Y}^n(t)=\hat{y}]= \frac1n(n-n\hat{y})\left[1-\sum_{k=1}^m \left(1-\frac{ np\hat{y}}{n-1}\right)^{n_k}p_k(t)\right]\\
&&\hspace{60pt}\approx \frac1n (n-n\hat{y})\frac{ np\hat{y}}{n-1}\E[N([nt])] \\
&&\hspace{60pt}\approx  p\hat{y}(1-\hat{y})\E[N([nt])]
=  \theta(t) \hat{y}(1-\hat{y})\Delta t
\end{eqnarray*}
using the approximation $1-(1-x)^a\approx ax$ for small $x$. With the same procedure we have
\begin{eqnarray*}
&&\hspace{-10pt}\textrm{Var}[\widehat{Y}^n(t+\Delta t)-\widehat{Y}^n(t)| \widehat{Y}^n(t)=\hat{y}]
=\frac{1}{n^2}\textrm{Var}\left[{Y}([nt]+1)-{Y}([nt])| {Y}([nt])=n\hat{y}\right]\\
&&\hspace{40pt}= \frac{1}{n^2}(n-n\hat{y}) \left[ 1 - \sum_{k=1}^m \left( 1-\frac{np\hat{y}}{n-1} \right)^{n_k} \,p_k(t) \right] \left[ \sum_{k=1}^m \left( 1-\frac{np\hat{y}}{n-1} \right)^{n_k} \,p_k(t) \right]\\
&&\hspace{40pt}\approx \frac1n(1-\hat{y})\left[1-\frac{np\hat{y}}{n-1}\E[N([nt])]\right] \frac{np\hat{y}}{n-1} \E[N([nt])]\\
&&\hspace{40pt}\approx \frac1n p (1-\hat{y})\hat{y} \E[N([nt])] = \frac{\theta(t)}{n}\hat{y}(1-\hat{y})\Delta t.
\end{eqnarray*}
Using as before the approximation $1-(1-x)^a\approx ax$ for small $x$ and $p(1-p)\approx p$ for small $p$.
As in the case, with recovery ($r<\infty$), following the same steps we can assume
\[\theta(t)=np\E[N([nt])] \approx npN:=\theta\]
for a constant $N$.

With this results and approximation methods for continuous time Markov chains using diffusion processes we can approximate $\widehat{Y}^n$ by a diffusion process $\widehat{Y}$ in $[0,1]$ that satisfies the stochastic differential equation
\begin{equation}\label{eq:dif1}
d\widehat{Y}(t)=\theta\widehat{Y}(t)(1-\widehat{Y}(t))dt+\sqrt{\frac{\theta}{n}\widehat{Y}(t)(1-\widehat{Y}(t))}dW(t),
\end{equation}
where $W=\{W(t),\,t\geq 0\}$ is a standard Brownian motion.
The approximation of diffusion processes by Markovian chains is well explained in \cite{Ibe2013} (Chapter 10).
\note{Another references: \cite{CerratoEA2011} (Section 3),  and \cite{VanKampen1982}.} 

If the variability is assumed to be small then the noise term of the stochastic differential equation \eqref{eq:dif1} will have little effect. In that case, we can presume that the mean of $\widehat{Y}(t)$ can be approximated by $\widehat{m}(t)$, where $\widehat{m}$ satisfies the deterministic equation
\[d\widehat{m}(t)=\theta\widehat{m}(t)(1-\widehat{m}(t))dt\]
with explicit solution
\[\widehat{m}(t)=\frac{1}{1+\frac{1-\widehat{m}_0}{\widehat{m}_0}e^{-\theta t}},\qquad t\geq 0,\]
where $\widehat{m}_0=\E[\widehat{Y}^n(0)]=\frac{1}{n}\E[Y(0)]$.

This suggest that for $t=0,1,2,\ldots$ (with attendant scaling up of error terms),
\[\E[Y(t)]=n\E\left[\widehat{Y}^n\left(\frac{t}{n}\right)\right]\approx n\widehat{m}\left(\frac{t}{n}\right)=\frac{n}{1+\frac{n-y_0}{y_0}e^{-pNt}},\]
where $y_0=\E[Y(0)]$.

\section{Simulations} \label{section.simulations}

The aim of this chapter is to simulate the SIR epidemic model we have presented in the previous sections, comparing the evolution of the disease for different distributions of the number of daily contacts $N(t)$. Particularly, we study the case when $N(t)$ follows a binomial distribution and compare the results for different values of its parameters.

As we are interested in showing the different evolution of the SIR epidemic model as the number of daily contacts changes, we fix the other parameters.
We consider a population of $n=10000$ individuals with a unique initial infective individual: $Y(0)=1$. The probability of transmission derived from an encounter is $p=0.1$ and the duration of the disease is $r=5$. We assume the quarantine is nonexistent, that is, $\lambda=1$.

The simulations are made using \texttt{R} programming environment. All of them are over $10000$ trials and run for a maximum time of 365 days.

\subsection{Binomial case}

Here, the number of daily contacts $N(t)$ has a binomial distribution with probability depending on time. This is $N(t)\sim Binom(N,p(t))$. With this distribution, the set of all possible values of $N(t)$ is $M=\{0,\dots,N\}$.

We consider three cases, in all of them the function $p(t)$ is periodic with yearly, monthly and weekly periods, respectively. In the first case, we fix a value for $N$ and compare the spread of the disease for four different functions of $p(t)$. In the other two cases, we choose a function for $p(t)$ and observe how the disease evolves when the maximum number of daily contacts changes.

\subsubsection*{Yearly period}

It is a natural assumption that the number of daily contacts varies all over the year and changes according to the season. We suppose that the mean of daily encounters arrives at its maximum in summer, decreases in fall until arriving at its minimum in winter, then increases in spring and summer again. Moreover, we assume that $p(t)$ has a sinusoidal shape, where the peak coincides with summer and the valley with winter, and we analyze the spread of the disease depending on the epoch of the year it starts. The nomenclature we use refers to this epoch.

The four functions we consider are:
\begin{eqnarray*}
	\textit{Spring:} \quad && p(t) = 0.3\cdot\sin\left(\frac{2\pi t}{365}\right)+0.5 \\
	\textit{Summer:} \quad && p(t) = 0.3\cdot\cos\left(\frac{2\pi t}{365}\right)+0.5 \\
	\textit{Fall:} \quad && p(t) = -0.3\cdot\sin\left(\frac{2\pi t}{365}\right)+0.5 \\
	\textit{Winter:} \quad && p(t) = -0.3\cdot\cos\left(\frac{2\pi t}{365}\right)+0.5
\end{eqnarray*}
All of them oscillate between 0.2 and 0.8. We fix the maximum number of daily contacts $N$ at 10.

\begin{figure}[h!]
	\centering
	\includegraphics[width=0.9\textwidth]{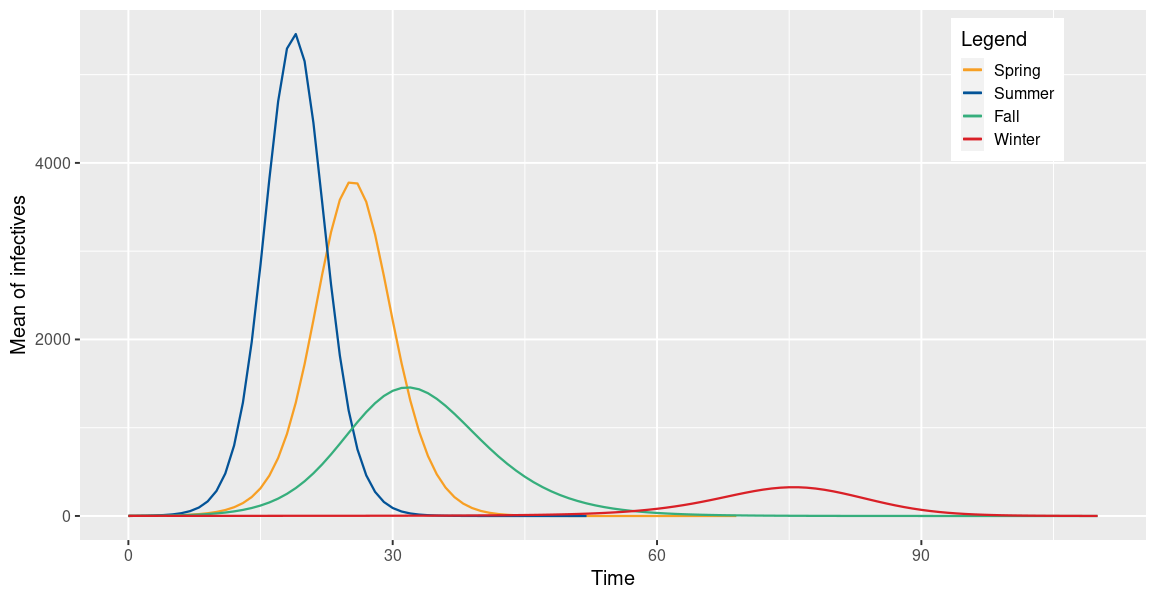}
	\begin{minipage}{.5\textwidth}
		\centering
		\vspace{5mm}
		\includegraphics[width=\textwidth]{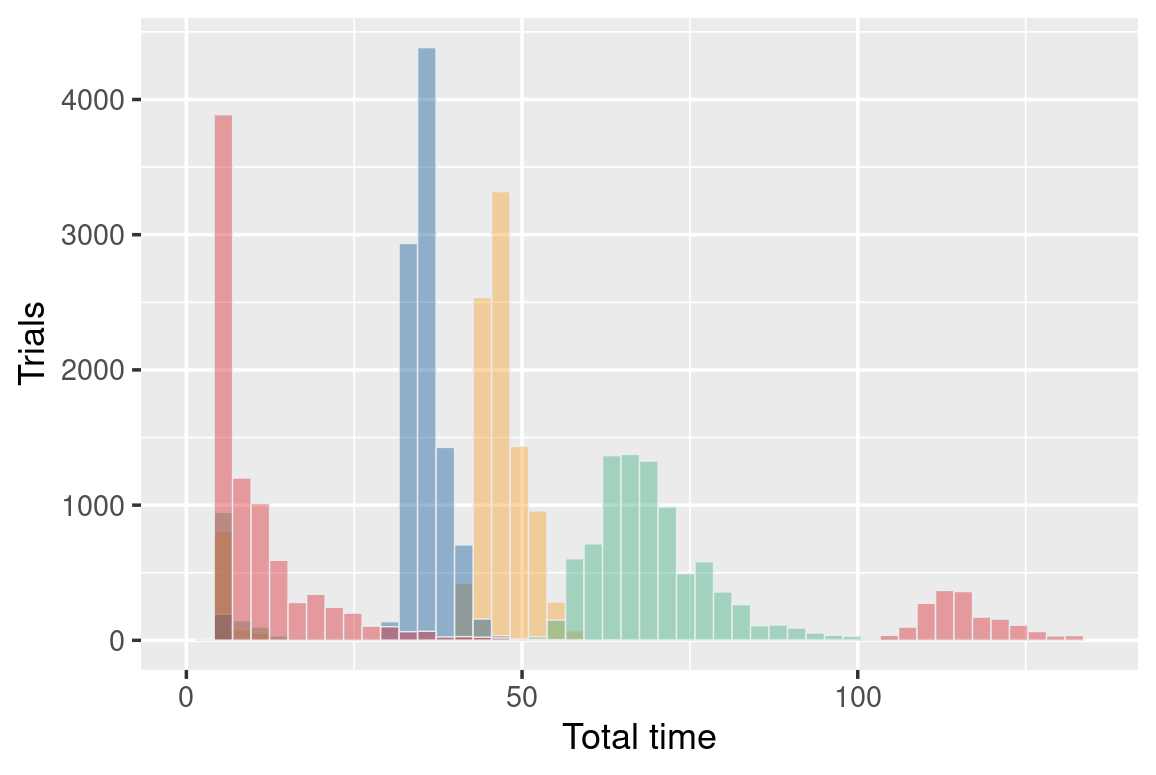}
	\end{minipage}%
	\begin{minipage}{.5\textwidth}
		\centering
		\vspace{5mm}
		\includegraphics[width=\linewidth]{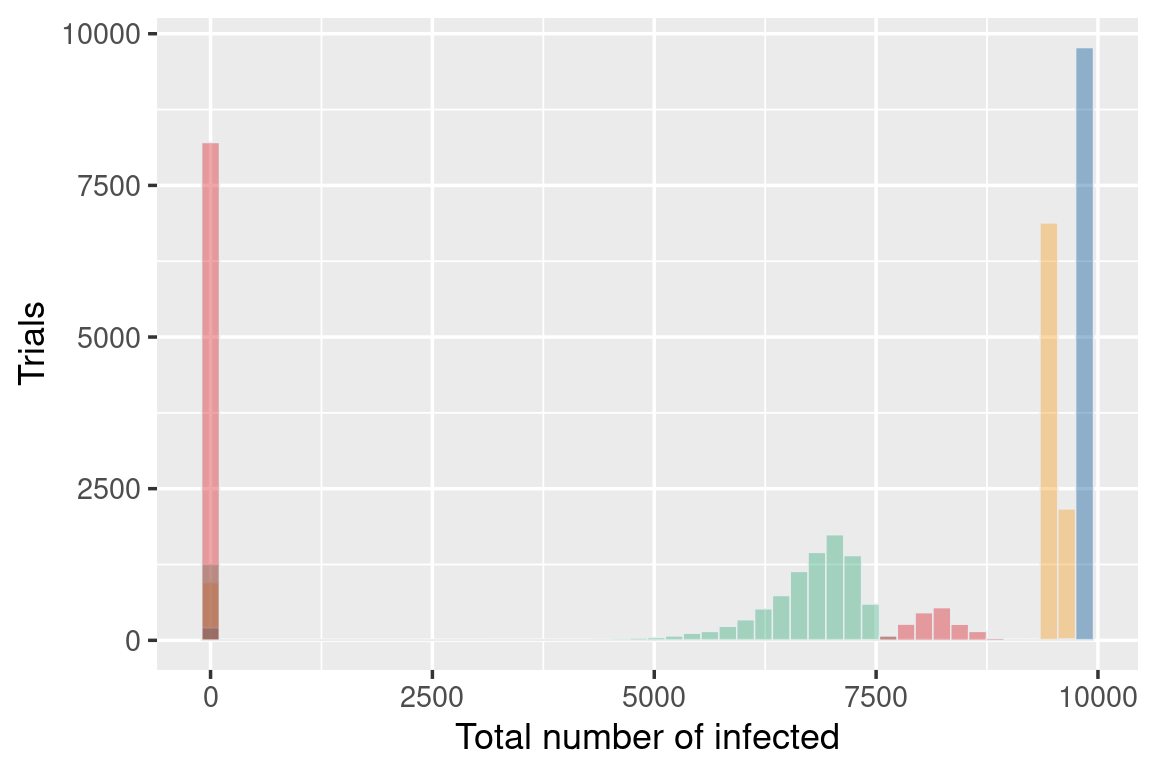}
	\end{minipage}
	\caption{SIR epidemic model depending on the season the disease starts. $p(t)$ is year-periodic and varies for each case, while $N$ is fixed to $N=10$.
		Top: Evolution of the mean number of infectives.
		Left: Histogram of the duration of the disease in days.
		Right: Histogram of the total number of infected in the population when the disease ends.}
	\label{fig.binomyear}
\end{figure}

Figure \ref{fig.binomyear} (Top) shows the evolution of the mean number of infectives per day. The graphic has been truncated at 110 days, as the number of cases of the following days is almost irrelevant.
We have a fast spread of the disease in the \textit{Summer} case, which coincides with the time when the number of encounters is higher. By reducing this number, the disease presents a lower mean number of infectives per day, but it lasts longer.
Observe that in \textit{Spring} and \textit{Fall} $p(t)$ has the same initial value $p(0)=0.5$, but the model evolves differently depending on whether this probability increases or decreases.

All these facts are underlined in Figure \ref{fig.binomyear} (Left and Right), that represent the histogram of the duration of the disease and the histogram of the total number of infected in the population when the disease ends.
Moreover, it is interesting to notice that, while in \textit{Spring} and \textit{Summer} almost the entire population gets infected in most trials, in \textit{Winter} the epidemic develops only in a limited number of cases.

\begin{table}
	\centering
	\def\arraystretch{1.5}
	\begin{tabular}{|c|c|c|c|c|}
		\hline
		& \textit{Spring} & \textit{Summer} & \textit{Fall} & \textit{Winter} \\
		\hline
		Total time & 43.3 (12.57) & 35.3 (5.10) & 61.5 (22.86) & 29.5 (41.75) \\
		\hline
		Population infected (\%) & 86\% (28\%) & 96\% (14\%) & 59\% (23\%) & 15\% (31\%) \\
		\hline
		$\widetilde{R_0}$ & 2.6 (1.27) & 4.0 (1.45) & 2.4 (1.25) & 1.0 (0.90) \\
		\hline
	\end{tabular}
	\caption{SIR epidemic model depending on the season the disease starts. $p(t)$ is year-periodic and varies for each case, while $N$ is fixed to $N=10$.
		Mean (standard deviation) of three variables: duration of the disease in days, total proportion of infecteds when the disease ends and number of individuals infected by the first infective ($\widetilde{R_0}$).}
	\label{table.binomyear}
\end{table}

Table \ref{table.binomyear} contains the average and the variation (mean and standard deviation) of the duration of the disease, the total proportion of the infected population when the disease ends, and the number of individuals infected by the first infective, which we denote by $\widetilde{R_0}$, for each case presented.

The first two rows help to better understand Figure \ref{fig.binomyear} (Left and Right).
The course of the disease is on average longer in \textit{Fall}, but it has a big variation, and it is on average shorter in \textit{Winter}, with an even bigger dispersion. This is evident in Figure \ref{fig.binomyear} (Left), where in \textit{Winter} we notice two different behaviors: in most of the trials the disease lasts for a short time and does not exceed 50 days, but in the other cases it lasts for more than 100 days.
As remarked before, almost the entire population is infected in \textit{Spring} and \textit{Summer}, but this proportion decreases drastically in \textit{Fall} and \textit{Winter}.

The last information contained in Table \ref{table.binomyear} is the number of individuals infected by the first infective.
In the first three cases, $\widetilde{R_0}$ exceeds the threshold value 1, implying that the disease turns into an epidemic. In the last case, the mean of $\widetilde{R_0}$ coincides with the threshold itself, but its variability makes unclear if an epidemic occurs on average or not.
Moreover, observe that $\widetilde{R_0}$ is very similar in \textit{Spring} and \textit{Fall}.

Finally, notice that the dispersion is smaller in \textit{Summer} and bigger in \textit{Winter} when we refer to the duration of the disease and the total proportion of infecteds, but it is reversed when we consider $\widetilde{R_0}$.

\subsubsection*{Monthly period}

We assume that the number of daily contacts changes periodically over the course of the month. This behavior could be due to economic reasons. For example, it might catch the tendency to go out more when a salary is perceived.
The periodicity is reflected in $p(t)$, that is the sinusoidal function
	\[ p(t) = 0.3\cdot\sin\left(\frac{2\pi t}{30}\right)+0.5. \]
This probability oscillates between 0.2 and 0.8.
We simulate and then compare the SIR epidemic model for three different values of the maximum number of daily contacts: $N=7$, $N=10$ and $N=12$.

\begin{figure}[h!]
	\centering
	\begin{minipage}{.5\textwidth}
		\centering
		\vspace{5mm}
		\includegraphics[width=\textwidth]{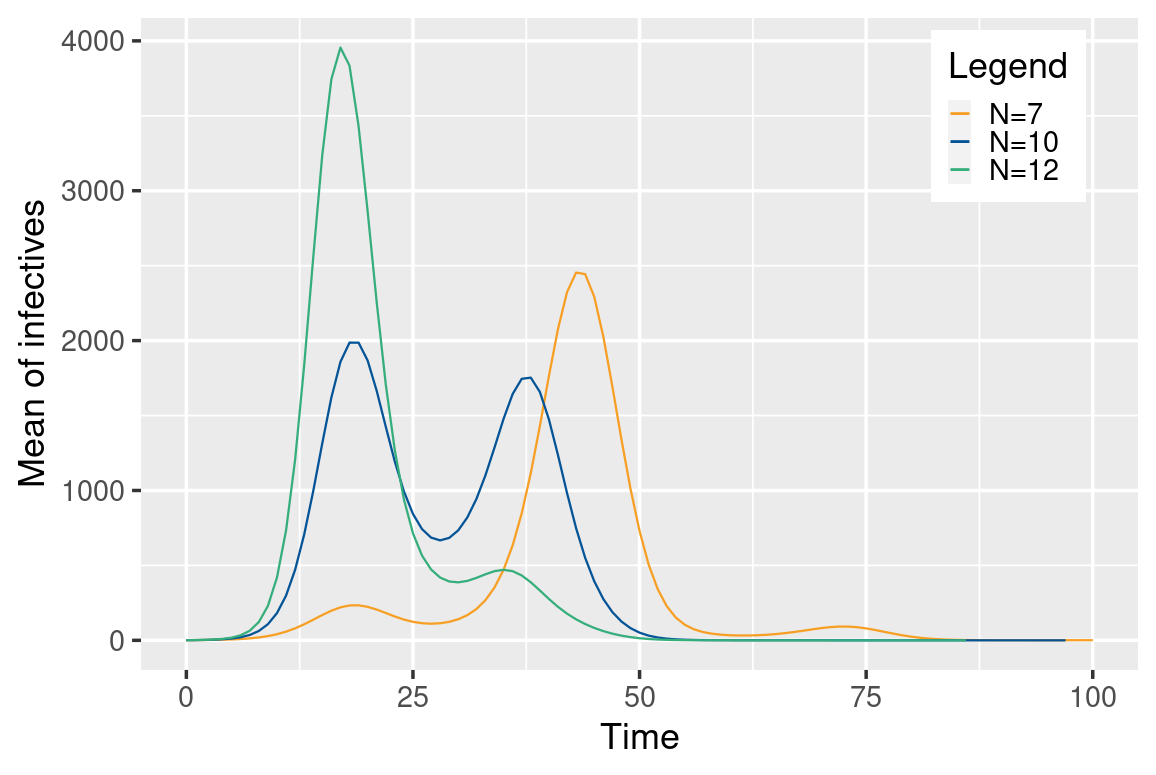}
	\end{minipage}%
	\begin{minipage}{.5\textwidth}
		\centering
		\vspace{5mm}
		\includegraphics[width=\linewidth]{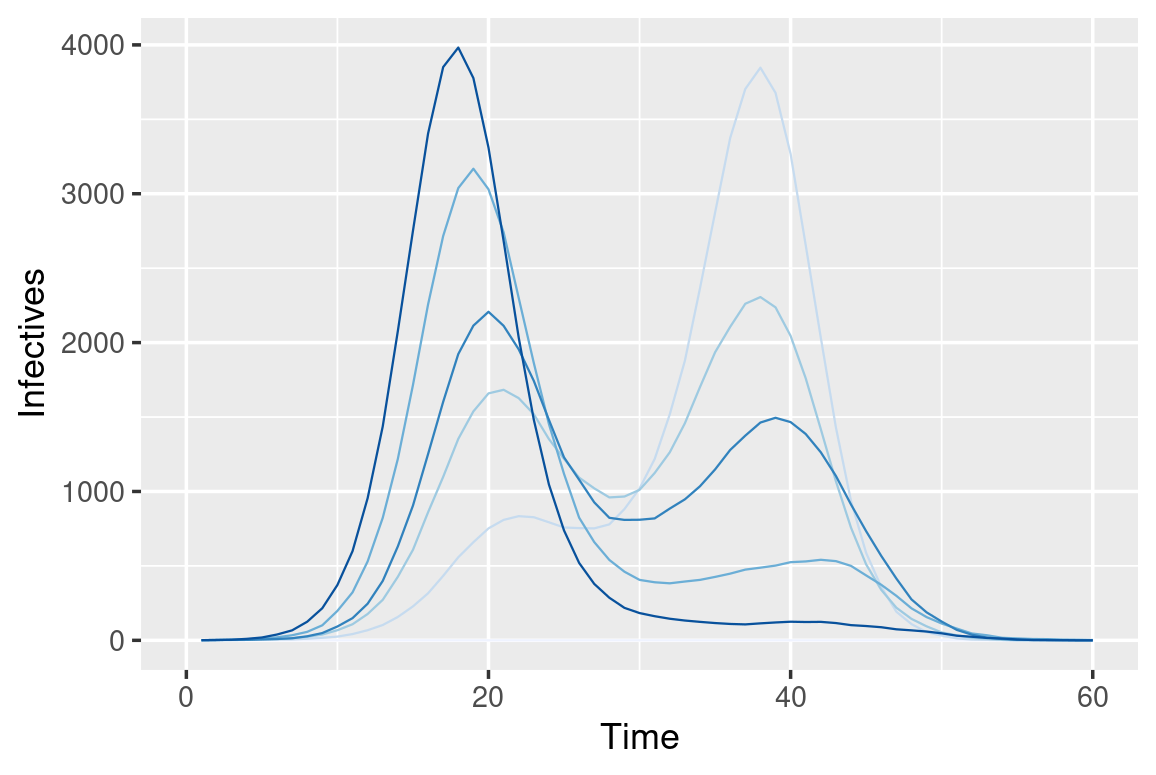}
	\end{minipage}
	\caption{SIR epidemic model with $p(t)$ monthly-periodic.
		Left: Evolution of the mean number of infectives for three different values of $N$.
		Right: Evolution of the number of infectives in some simulations for $N=10$.}
	\label{fig.binommonth}
\end{figure}

Figure \ref{fig.binommonth} (Left) shows the evolution of the mean number of infectives per day in the three cases.
The graphic has been truncated at 100 days.
We see that the evolution of the disease behaves differently with respect to the maximum number of daily encounters. Particularly, the number and position of the peaks vary.
When $N=7$, there are three peaks, the central one big enough and the others very small.
When the maximum number of daily contacts is set to $N=10$, two similar peaks are displayed.
Finally, when this parameter is increased to $N=12$, a tall peak is followed by a small one.

It might be interesting to investigate how the mean number of infectives presents two peaks in the case $N=10$.
Figure \ref{fig.binommonth} (Right) shows the evolution of the number of infectives per day for some selected simulations. Observe that the trials behave quite differently: in some cases there are two (similar or not) peaks, in others only one. However, it is also important to point out that all peaks occur at the same time as those of the mean.

\begin{figure}[h!]
	\centering
	\begin{minipage}{.5\textwidth}
		\centering
		\vspace{5mm}
		\includegraphics[width=\textwidth]{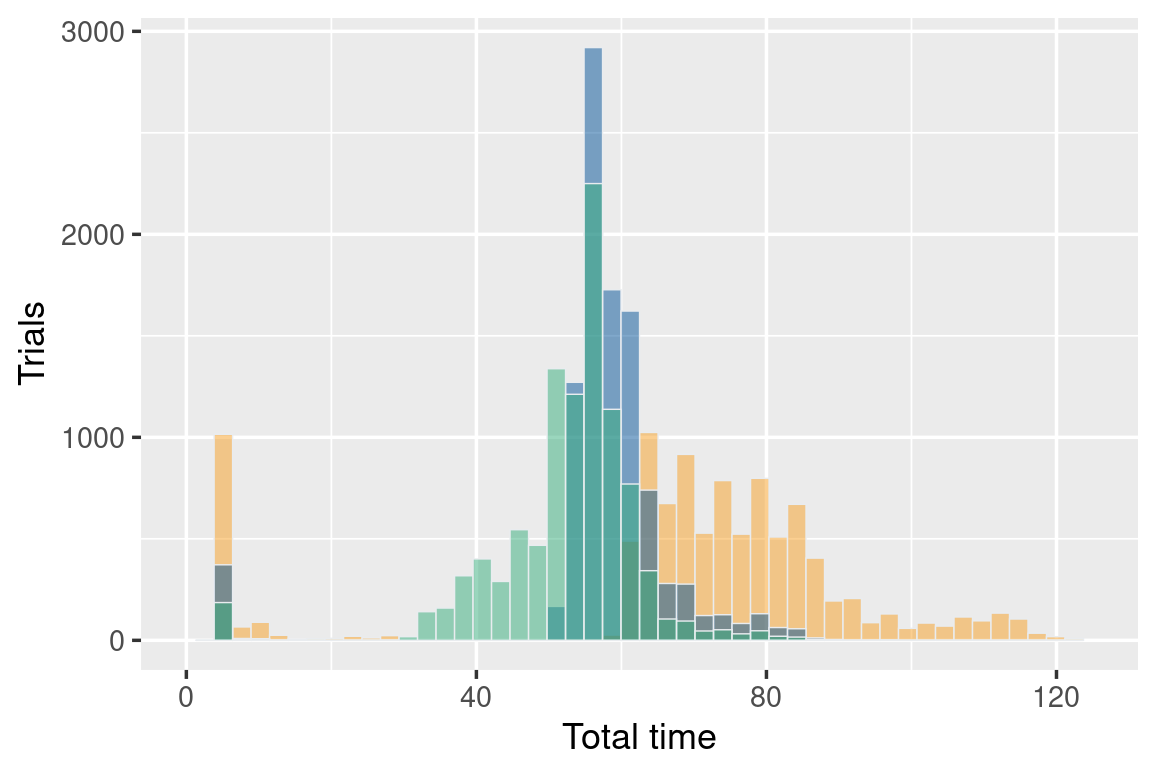}
	\end{minipage}%
	\begin{minipage}{.5\textwidth}
		\centering
		\vspace{5mm}
		\includegraphics[width=\linewidth]{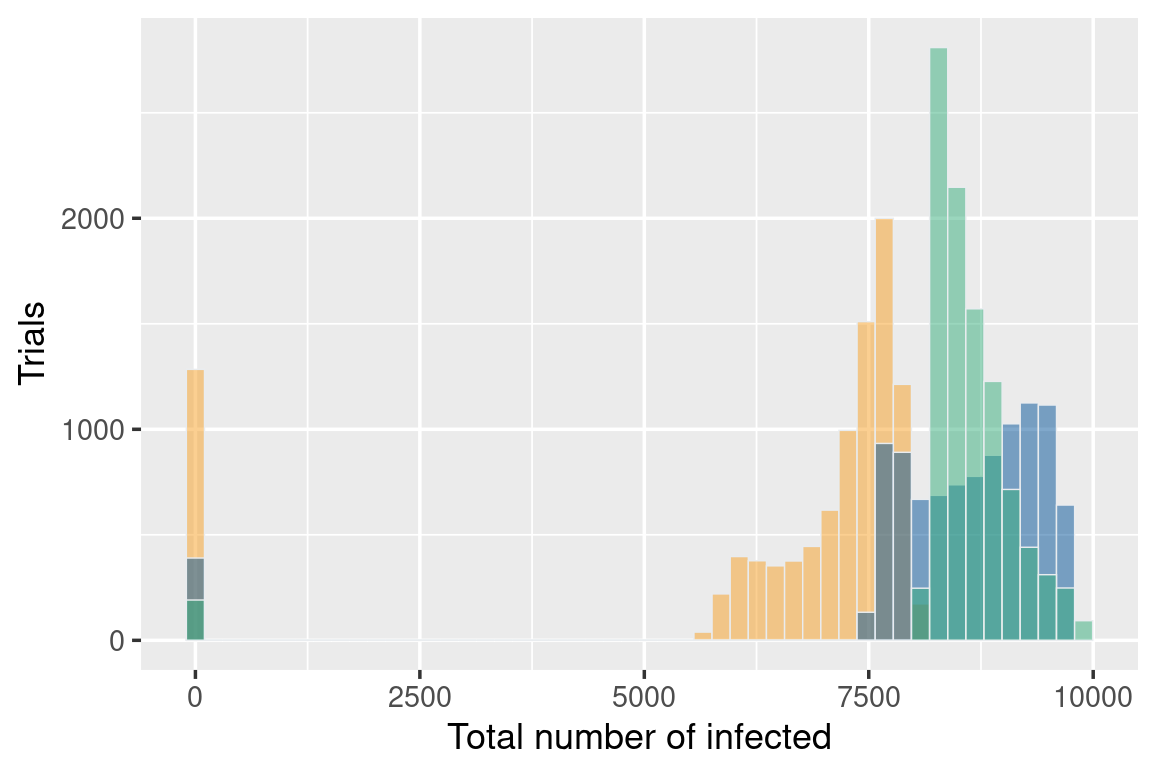}
	\end{minipage}
	\caption{SIR epidemic model with $p(t)$ monthly-periodic. Comparison for $N=7$, $N=10$ and $N=12$ (same colors of Figure \ref{fig.binommonth}).
		Left: Histogram of the duration of the disease in days.
		Right: Histogram of the total number of infected in the population when the disease ends.}
	\label{fig.binommonth.hist}
\end{figure}

It appears that on average the disease lasts longer when the number of daily contacts is smaller. This is noted in Figure \ref{fig.binommonth.hist} (Left) and Table \ref{table.binommonth}.
In addition, here we observe a bigger variation for smaller values of $N$. See, for example, that the standard deviation of the case $N=7$ is more than double that of the case $N=10$.
This is also due to the fact that the disease does not spread in more simulations of the case $N=7$ than of the others, as pointed out in Figure \ref{fig.binommonth.hist} (Left).

Figure \ref{fig.binommonth.hist} (Right) shows the distribution of the total number of infected over all simulations.
When the number of contacts increases, this quantity is on average bigger but its variation decreases, as indicated in Table \ref{table.binommonth}.
Note that the cases in which the total population is infected are very few.

\begin{table}
	\centering
	\def\arraystretch{1.5}
	\begin{tabular}{|c|c|c|c|}
		\hline
		& $N=7$ & $N=10$ & $N=12$ \\
		\hline
		Total time & 68.5 (26.76) & 57.4 (12.06) & 52.6 (10.28) \\
		\hline
		Population infected (\%) & 63\% (25\%) & 84\% (18\%) & 85\% (12\%) \\
		\hline
		$\widetilde{R_0}$ & 2.4 (1.25) & 3.3 (1.37) & 4.0 (1.44) \\
		\hline
	\end{tabular}
	\caption{SIR epidemic model with $p(t)$ monthly-periodic and three different values of $N$.
		Mean (standard deviation) of three variables: duration of the disease in days, total proportion of infecteds when the disease ends and number of individuals infected by the first infective ($\widetilde{R_0}$).}
	\label{table.binommonth}
\end{table}

Finally, Table \ref{table.binommonth} reveals that the mean number of individuals infected by the first infective increases as the maximum number of daily encounters does. Its variation follows the same tendency, although the differences are quite small.

\subsubsection*{Weekly period}

Now, we assume that the number of daily contacts changes periodically with a week period. For example, this could mark the different behavior of the population at the weekend with respect to the weekdays.
The periodicity is reflected in $p(t)$, that is the sinusoidal function
	\[ p(t) = 0.3\cdot\cos\left(\frac{2\pi t}{7}\right)+0.5. \]
This probability oscillates between 0.2 and 0.8.
As in the monthly-periodic study, we simulate and then compare the SIR epidemic model for three different values of the maximum number of daily contacts: $N=7$, $N=10$ and $N=12$.

\begin{figure}[h!]
	\centering
	\begin{minipage}{.5\textwidth}
		\centering
		\vspace{5mm}
		\includegraphics[width=\textwidth]{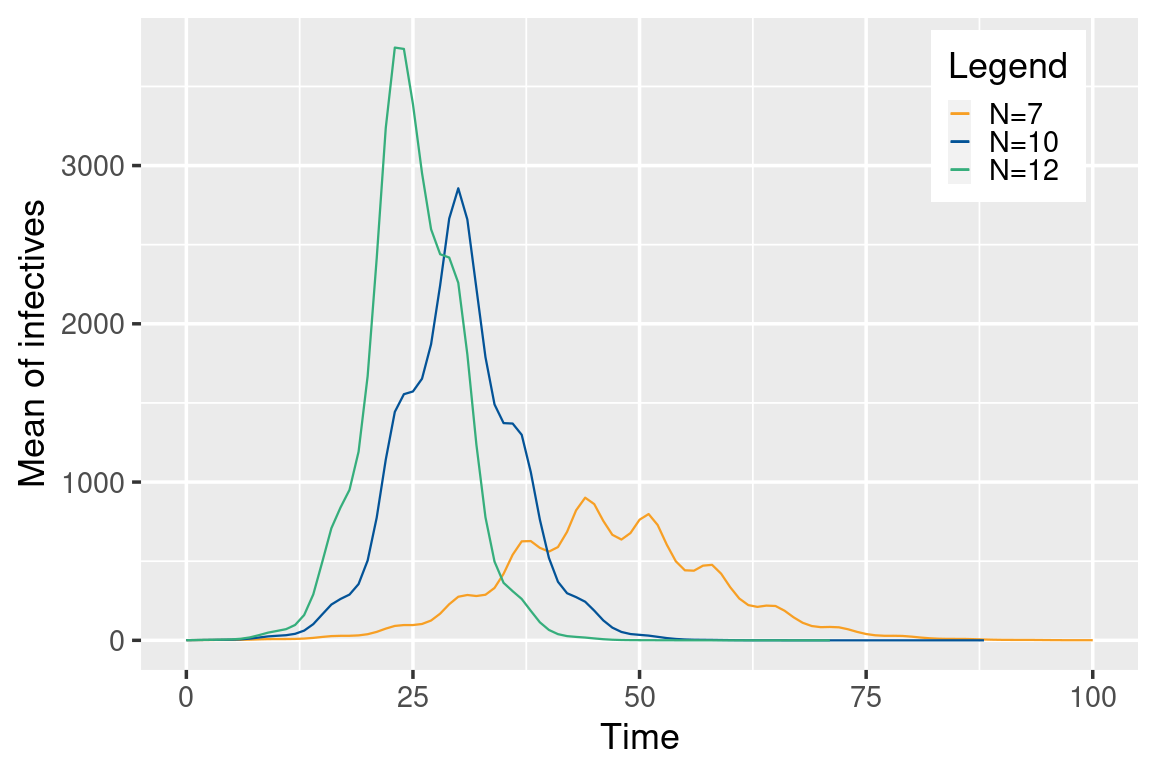}
	\end{minipage}%
	\begin{minipage}{.5\textwidth}
		\centering
		\vspace{5mm}
		\includegraphics[width=\linewidth]{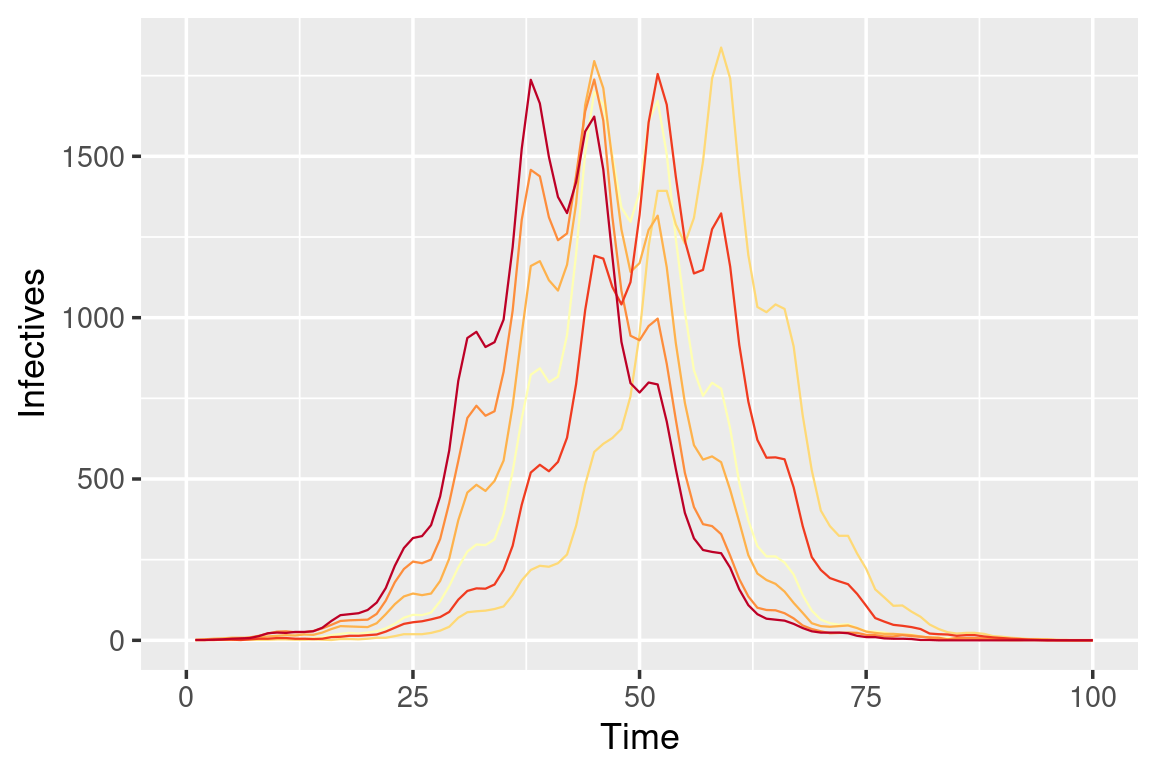}
	\end{minipage}
	\caption{SIR epidemic model with $p(t)$ weekly-periodic.
		Left: Evolution of the mean number of infectives for three different values of $N$.
		Right: Evolution of the number of infectives in some simulations for $N=7$.}
	\label{fig.binomweek}
\end{figure}

Figure \ref{fig.binomweek} (Left) plots the evolution of the mean number of infectives in the three cases.
The graphic has been truncated at 100 days.
One could observe that the results are very different depending on the values of $N$.
When $N=7$, the mean number of infectives oscillates many times, presenting several peaks, but none of them is particularly high.
Some selected simulations are displayed in Figure \ref{fig.binomweek} (Right). They seem to reflect the same tendency of the mean.
When $N=10$ and $N=12$, only one peak is displayed, but the curves are not similar to those of a normal distribution, as they were for the yearly and monthly periodic studies. Furthermore, the different shapes the two of them presented make their behavior quite distinct.

\begin{figure}[h!]
	\centering
	\begin{minipage}{.5\textwidth}
		\centering
		\vspace{5mm}
		\includegraphics[width=\textwidth]{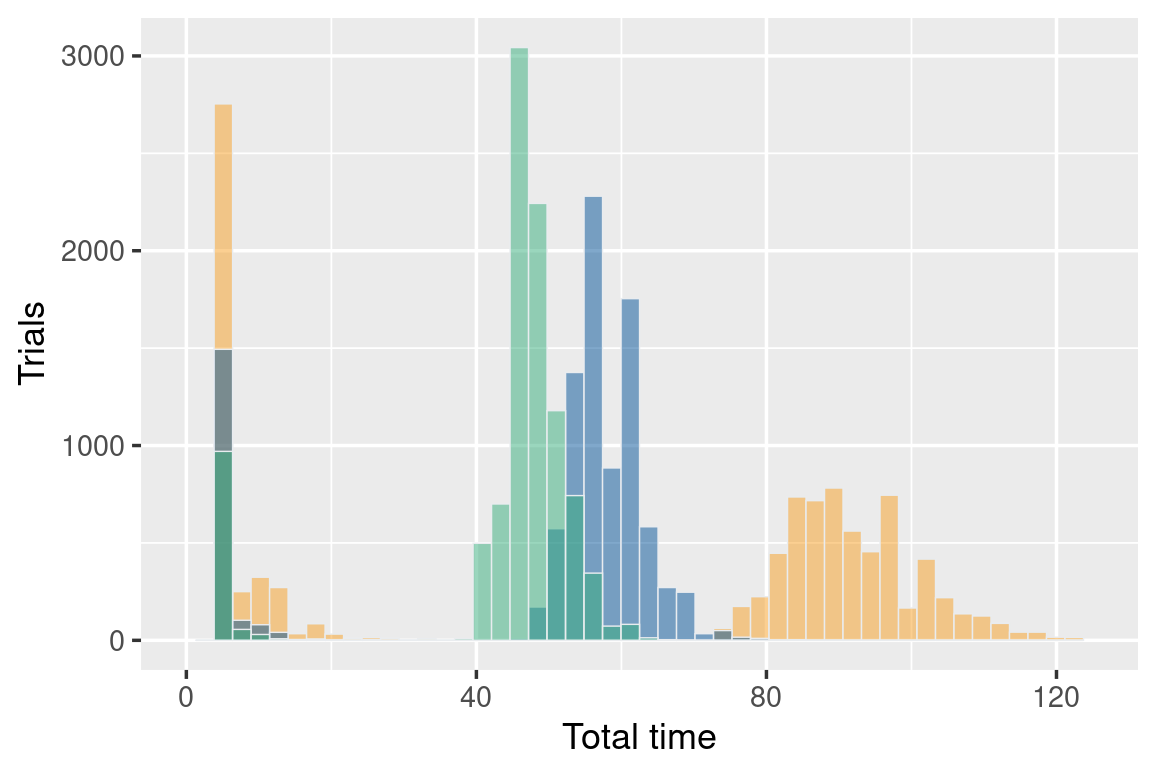}
	\end{minipage}%
	\begin{minipage}{.5\textwidth}
		\centering
		\vspace{5mm}
		\includegraphics[width=\linewidth]{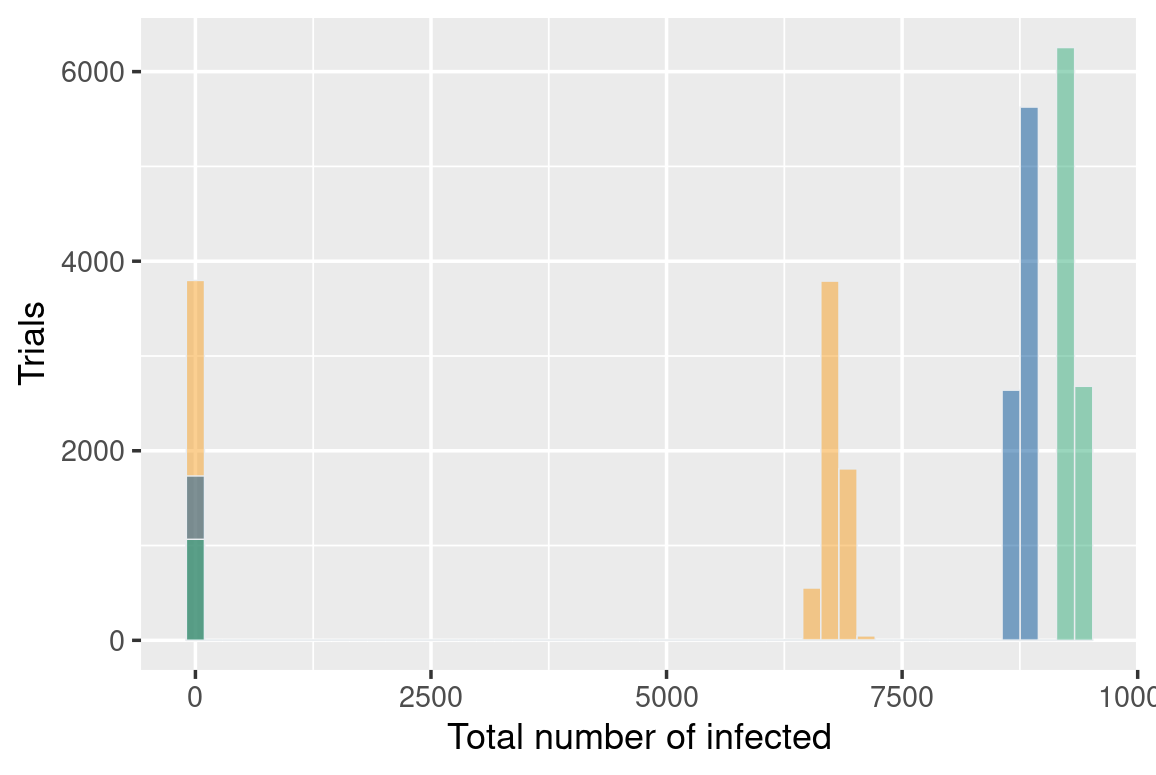}
	\end{minipage}
	\caption{SIR epidemic model with $p(t)$ weekly-periodic. Comparison for $N=7$, $N=10$ and $N=12$ (same colors of Figure \ref{fig.binomweek}).
		Left: Histogram of the duration of the disease in days.
		Right: Histogram of the total number of infected in the population when the disease ends.}
	\label{fig.binomweek.hist}
\end{figure}

The histograms in Figure \ref{fig.binomweek.hist} (Left and Right) also highlight the differences that the change in the number of daily contacts generates.
We observe two features also reported in the monthly-periodic study. On one hand, the disease lasts longer on average when the number of daily contacts is smaller. On the other hand, the magnitude of the spread, intended as the part of the total population infected, decreases.
Furthermore, note that when $N=7$ the number of trials in which the disease does not spread is much higher than in the other two cases.
This increases the variability of the results, as it can be observed in Table \ref{table.binomweek}.

\begin{table}
	\centering
	\def\arraystretch{1.5}
	\begin{tabular}{|c|c|c|c|}
		\hline
		& $N=7$ & $N=10$ & $N=12$ \\
		\hline
		Total time & 59.7 (41.94) & 48.9 (20.25) & 43.6 (13.68) \\
		\hline
		Population infected (\%) & 42\% (33\%) & 73\% (33\%) & 83\% (29\%) \\
		\hline
		$\widetilde{R_0}$ & 1.4 (1.05) & 2.0 (1.20) & 2.4 (1.28) \\
		\hline
	\end{tabular}
	\caption{SIR epidemic model with $p(t)$ weekly-periodic and three different values of $N$.
		Mean (standard deviation) of three variables: duration of the disease in days, total proportion of infecteds when the disease ends and number of individuals infected by the first infective ($\widetilde{R_0}$).}
	\label{table.binomweek}
\end{table}

The dispersion of the mean number of individuals infected by the first infective is also very high in all cases (see Table \ref{table.binomweek}). Particularly, when $N=7$ and $N=10$, it is difficult to know if the spread of the disease will result in an epidemic or not.

\section{Conclusions}

In this paper, we study a stochastic SIR-type epidemic model that is an extension of the one proposed by Tuckwell and Williams. We assume that the number of daily encounters of each individual depends on time and we add a parameter to control a possible quarantine of the infectious individuals. Two cases are taken into consideration: when the duration of the disease is constant and when infectious individuals remain infectious throughout their life.
In both situations, we describe analytically the underlying model and its dynamics, deriving a diffusion process and the basic reproduction number. 
Several simulations are made to show how differently the disease evolves with respect to the distribution of the number of daily encounters. The dependence on time of this parameter also plays an important role, as models that begin with the same parameters in the first epoch can evolve in very different ways.

These results can be the springboard to construct more complex models to investigate diseases that cannot be adequately described by only the three classes S, I and R. For example, diseases with an initial latency period or the presence of asymptomatic individuals.
In these cases, it could be interesting to explore how different distributions and the dependence on time of some parameters, such as the number of daily encounters, can affect the evolution of the disease.

\bibliographystyle{alpha} 
\bibliography{Biblio_Giu}

\end{document}